\documentclass[11pt]{article}

\usepackage{indentfirst}
\usepackage[margin=1.1in]{geometry}
\usepackage{amsmath}
\usepackage{multirow}
\usepackage{graphicx}
\usepackage{multirow}
\usepackage{authblk}
\usepackage{apacite}

\title{Non-Extensive Value-at-Risk Estimation During Times of Crisis}
\author[1]{Ahmad Hajihasani \thanks{Email: a.hajihasani@ut.ac.ir}}
\author[1,2]{Ali Namaki \thanks{Corresponding author, Email: alinamaki@ut.ac.ir}}
\author[1]{Nazanin Asadi \thanks{Email: n.asadi@ut.ac.ir}}
\author[1]{Reza Tehrani \thanks{Email: rtehrani@ut.ac.ir}}
\affil[1]{Department of Finance, Faculty of Management, University of Tehran, Tehran, Iran}
\affil[2]{Iran Finance Association, Tehran, Iran}

\date{}

\begin{document}

\maketitle

\abstract
Value-at-risk is one of the important subjects that extensively used by researchers and practitioners for measuring and managing uncertainty in financial markets. Although value-at-risk is a common risk control instrument, but there are criticisms about its performance. One of these cases, which has been studied in this research, is the value-at-risk underestimation during times of crisis. In these periods, the non-Gaussian behavior of markets intensifies and the estimated value-at-risks by normal models are lower than the real values. In fact, during times of crisis, the probability density of extreme values in financial return series increases and this heavy-tailed behavior of return series reduces the accuracy of the normal value-at-risk estimation models. A potential approach that can be used to describe non-Gaussian behavior of return series, is Tsallis entropy framework and non-extensive statistical methods.\par
In this paper, we have used non-extensive value at risk model for analyzing the behavior of financial markets during times of crisis. By applying q-Gaussian probability density function, we can see a better value-at-risk estimation in comparison with the normal models, especially during times of crisis. We showed that q-Gaussian model estimates value-at-risk better than normal model. Also we saw in the mature markets, it is obvious that the difference of value-at-risk between normal condition and non-extensive approach increase more than one standard deviation during times of crisis, but in the emerging markets we cannot see a specific pattern.

\section{Introduction}
Financial markets can be seen as complex networks of heterogeneous agents which are coupled with each other and with their environments \cite{ARTICLE:31}\cite{ARTICLE:32}\cite{ARTICLE:33}\cite{ARTICLE:30}\cite{ARTICLE:35}\cite{ARTICLE:36}. The result of interaction between these agents reflects in prices. Financial studies, in most of the cases consider the price time-series as a simple Brownian motion which means return time-series has Gaussian distribution. Recent studies call this Gaussian assumption into question and reveal some deviations such as fat tails in financial market data. these anomalies are more significant during times of financial crisis\cite{10.1371/journal.pone.0160363}\cite{ARTICLE:14}\cite{ARTICLE:24}\cite{ARTICLE:6}\cite{ARTICLE:7}. A potential approach which addresses the concerns with non-Gaussian and multi-fractal properties of real market data is non-extensive statistical framework and Tsallis entropy \cite{ARTICLE:8}\cite{ARTICLE:26}\cite{ARTICLE:29}. This approach leads us to q-Gaussian probability density function in which q parameter is considered as quantization factor of non-Gaussian properties of the stochastic variable\cite{ARTICLE:18}\cite{ARTICLE:19}\cite{ARTICLE:21}\cite{ARTICLE:22}\cite{ARTICLE:34}.\par
Accurate risk assessment during financial crisis periods is a critical and crucial task. There are variety of models and methods to estimate financial risk. An early model is Marcowitz's model which assumes risk as standard deviation of return series \cite{ARTICLE:1}. Other risk measures consist of semi-variance\cite{ARTICLE:2}, Lower Partial Moment\cite{ARTICLE:3}, expected shortfall\cite{ARTICLE:4} and etc. However, one of the most practical instruments for risk management is value-at-risk. Value-at-risk evaluates maximum expected loss across specific time period \cite{BOOK:1}\cite{BOOK:2}\cite{ARTICLE:5}. In practice often return series have leptokurtic behaviors that cannot be described by Gaussian distribution \cite{ARTICLE:15}. This issue leads to underestimation of value at risk based on the normal conditions \cite{ARTICLE:16}. Applying entropy functions such as q-Gaussian distribution to estimate value-at-risk results more accurate measures due to non-Gaussian properties of financial time-series \cite{ARTICLE:23}\cite{ARTICLE:24}\cite{ARTICLE:25}.\par
Recently Tsallis entropy has been used by researchers to study the financial markets and their behavior during times of crisis. Gençay and Gradojevic compared the 1987 and 2008 financial crisis using entropic risk management measures and suggested the entropic methodology as a market sentiment indicator \cite{ARTICLE:9}. Bill et al. used non-extensive statistical framework to study the behavior of volatility in Polish stock market \cite{ARTICLE:10}. Borland have used Tsallis entropy framework to create an option pricing model \cite{ARTICLE:11}\cite{ARTICLE:27}\cite{ARTICLE:28}. Kozaki and Sato applied the non-extensive statistics in portfolio risk management \cite{ARTICLE:12}. Zhao et al. proposed an optimal portfolio selection model based on Tsallis non-extensive statistical mechanics using value-at-risk as constraint for optimization model \cite{ARTICLE:13}. Namaki et al. used non-extensive statistical framework for analyzing the financial markets during times of crisis and found that q parameter increases in these periods \cite{ARTICLE:14}.\par
In this paper, we have estimated q parameter based on return time-series of several mature and emerging markets; then we have compared the difference betwean normal and non-extensive value-at-risk in these markets over a twenty years time period. In addition, value-at-risk in various time scales and market conditions has been studied.
The empirical data for this research collected from mature and emerging markets from January 20, 2000 to March 20, 2019, except Tehran Stock Exchange price index data, which is available from January 20, 2009 to March 20, 2019. Dow Jones Industrial Average (DJIA), Tokyo Stock Exchange (Nikkei 225) and Frankfurt Stock Exchange (DAX) are selected as mature markets. Indices of Tehran Stock Exchange (TSE), Shanghai Stock Exchange (SSE) and Bombay Stock Exchange (BSE) are considered as emerging markets.\par
This research is organized in 4 sections. In part 2 the statistical models and methods will be defined. Data analysis and implementation of statistical models are considered in section 3. Section 4 is dedicated to conclusions.\par

\section{Methods}
A typical approach to express disorder and uncertainty in financial market data is to use the concept of entropy. The Shannon information theory formulates the entropy of a continuous random variable as,
\begin{equation}
S = -\int P(x)\ln(x)dx.
\end{equation}
In order to capture anomalous properties of financial time-series, such as fat tails and extreme values, Tsallis entropy has been used in this study \cite{ARTICLE:25}\cite{ARTICLE:19}\cite{ARTICLE:21}. Tsallis entropy function as a generalization of Boltzman-Gibbs entropy \cite{ARTICLE:17}\cite{ARTICLE:20}, is defined as,\par
\begin{equation}
S_q = \frac{1 - \int P(x)^q dx}{q-1}.
\end{equation}
In this framework the parameter q acts as an index for non-Gaussianity.\par
Under the following conditions,\par
\begin{align}
\int P(x) dx &= 1 \\
\int P^{(q)}(x) x^2 dx &= {\sigma_q}^2
\end{align}
where $P^{(q)}(x) = \frac{P^q(x)}{\int P^q(x)dx}$, The maximum entropy principle leads us to q-Gaussian probability density function as,\par
\begin{equation}
P_q(x) = \frac{1}{Z_q} \left( 1 - \frac{1-q}{3-q} \frac{x^2}{{\sigma_q}^2} \right)_+^\frac{1}{1-q}.
\end{equation}
In this equation $q \ne 1$, $q < 3$, $a_+ = max(a, 0)$ and $Z_q$ is normalization factor that can be obtained as below,\par
\begin{equation}
Z_q = \int \left( 1 - \frac{1-q}{3-q} \frac{x^2}{{\sigma_q}^2} \right)_+^\frac{1}{1-q} dx.
\end{equation}
Assuming $1 < q < 3$ then $Z_q = \left( \frac{3-q}{q-1} {\sigma_q}^2 \right)^\frac{1}{2} B(\frac{3-q}{2(q-1)}, \frac{1}{2})$.\par
Here $B(a, b) = \int t^{a-1} (1-t)^{b-1} dt$, $(a, b > 0)$ is the beta function.\par
Figure 1.a shows that as $q$ value gets closer to 1, q-Gaussian probability density function becomes more similar to normal probability density function.\par

\subsection{Non-Extensive Value-at-Risk (q-VaR)}
Value-at-risk can be defined as maximum expected loss over a time interval \cite{BOOK:1}\cite{BOOK:2}. Using q-Gaussian probability distribution function, the non-extensive form of value-at-risk can be represented as,\par
\begin{equation}
Pr(P_q(T)-P_q(0)<-VaR_q) = 1 - \alpha
\end{equation}
Where $T$ is time horizon, $\alpha$ is the confidence level and $P_q(x)$ is q-Gaussian probability density function.\par
Figure 1.b depicts the value-at-risk concept on a Gaussian probability density function diagram.\par

\begin{figure}
	\centering
	\begin{tabular}{@{}c@{}}
		\includegraphics[width=0.45\linewidth]{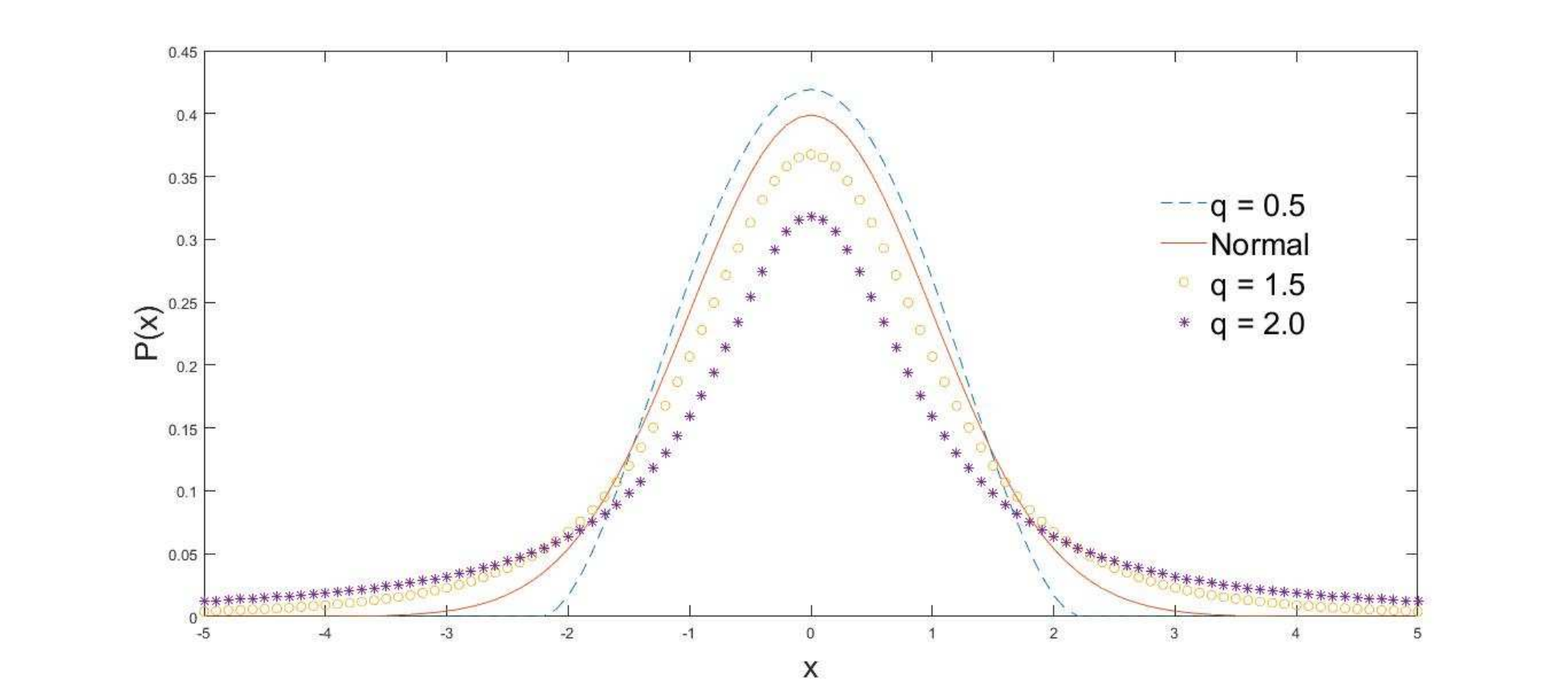}\\[\abovecaptionskip]
		\small (a)
	\end{tabular}
	\begin{tabular}{@{}c@{}}
		\includegraphics[width=0.45\linewidth]{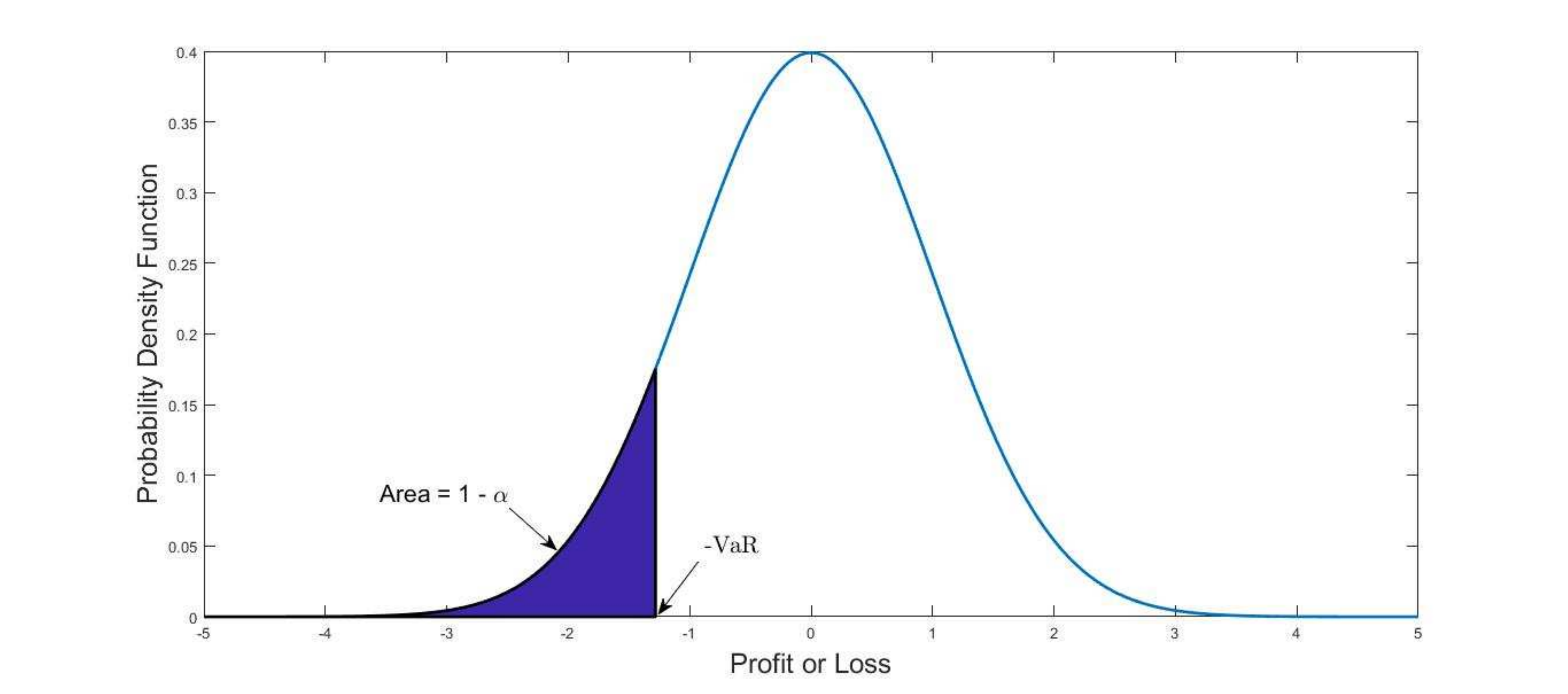}\\[\abovecaptionskip]
		\small (b)
	\end{tabular}
	\caption{\small (a) Standard q-Gaussian PDFs for different q parameters; (b) Value at Risk (The specified area equals to $1-\alpha$, where $\alpha$ is confidence level.)}
\end{figure}

To implement the statistical models and analysis of the results, we calculate daily return time series of price indices\par
\begin{equation}
r_t = ln \frac{x_t}{x_{t-1}},
\end{equation}
assuming $x_t$ is the price index value for t-th day. Then we normalize the resulted values as,\par
\begin{equation}
\bar{r_t} = \frac{r_t - \mu_r}{\sigma_r},
\end{equation}
where $\mu_r$ and $\sigma_r$ are the mean and standard deviation of return time series.

\begin{table}\footnotesize
	\begin{center}
		\caption{Back-test results for mature markets (n = number of observations)}
		\begin{tabular}{|c|c|c|c|c|c|c|c|c|c|c|}
			\hline
			\multirow{2}{*}{\textbf{$\alpha$}} & \multirow{2}{*}{\textbf{Method}} & \multicolumn{3}{c|}{\textbf{DJIA \tiny($q = 1.21\pm0.02, n = 4567$)}} & \multicolumn{3}{c|}{\textbf{N225 \tiny($q = 1.18\pm0.02, n = 4456$)}} & \multicolumn{3}{c|}{\textbf{DAX \tiny($q = 1.19\pm0.02, n = 4609$)}}\\
			\cline{3-11}
			& & VaR \tiny (percent) & \multicolumn{2}{c|}{Violations} & VaR \tiny (percent) & \multicolumn{2}{c|}{Violations} & VaR \tiny (percent) & \multicolumn{2}{c|}{Violations} \\ \hline
			\multirow{2}{*}{0.95} & q-Gaussian & 2.06 & 163 & \%3.57 & 2.71 & 155 & \%3.48 & 2.68 & 170 & \%3.69 \\
			& Gaussian & 1.83 & 211 & \%4.62 & 2.47 & 195 & \%4.38 & 2.41 & 222 & \%4.82 \\ \hline
			\multirow{2}{*}{0.96} & q-Gaussian & 2.22 & 128 & \%2.80 & 2.91 & 129 & \%2.89 & 2.88 & 139 & \%3.02 \\
			& Gaussian & 1.95 & 186 & \%4.07 & 2.63 & 166 & \%3.73 & 2.57 & 184 & \%3.99 \\ \hline
			\multirow{2}{*}{0.97} & q-Gaussian & 2.42 & 102 & \%2.23 & 3.17 & 98 & \%2.20 & 3.14 & 114 & \%2.47 \\
			& Gaussian & 2.10 & 155 & \%3.39 & 2.82 & 141 & \%3.16 & 2.76 & 158 & \%3.43 \\ \hline
			\multirow{2}{*}{0.98} & q-Gaussian & 2.71 & 73 & \%1.60 & 3.53 & 74 & \%1.66 & 3.50 & 82 & \%1.78 \\
			& Gaussian & 2.29 & 117 & \%2.56 & 3.09 & 110 & \%2.47 & 3.01 & 125 & \%2.71 \\ \hline
		\end{tabular}
	\end{center}
\end{table}

\begin{table}\footnotesize
	\begin{center}
		\caption{Back-test results for emerging markets (n = number of observations)}
		\begin{tabular}[width=\linewidth]{|c|c|c|c|c|c|c|c|c|c|c|}
			\hline
			\multirow{2}{*}{\textbf{$\alpha$}} & \multirow{2}{*}{\textbf{Method}} & \multicolumn{3}{c|}{\textbf{TSE \tiny($q = 1.22\pm0.02, n = 2455$)}} & \multicolumn{3}{c|}{\textbf{SSE \tiny($q = 1.21\pm0.02, n = 4501$)}} & \multicolumn{3}{c|}{\textbf{BSE \tiny($q = 1.20\pm0.02, n = 4477$)}}\\
			\cline{3-11}
			& & VaR \tiny (percent) & \multicolumn{2}{c|}{Violations} & VaR \tiny (percent) & \multicolumn{2}{c|}{Violations} & VaR \tiny (percent) & \multicolumn{2}{c|}{Violations} \\ \hline
			\multirow{2}{*}{0.95} & q-Gaussian & 1.34 & 70 & \%2.85 & 2.89 & 165 & \%3.67 & 2.52 & 146 & \%3.26 \\
			& Gaussian & 1.17 & 91 & \%3.70 & 2.57 & 207 & \%4.60 & 2.27 & 195 & \%4.36 \\ \hline
			\multirow{2}{*}{0.96} & q-Gaussian & 1.45 & 57 & \%2.32 & 3.11 & 133 & \%2.95 & 2.72 & 132 & \%2.95 \\
			& Gaussian & 1.26 & 80 & \%3.26 & 2.74 & 183 & \%4.07 & 2.41 & 161 & \%3.60 \\ \hline
			\multirow{2}{*}{0.97} & q-Gaussian & 1.60 & 44 & \%1.79 & 3.39 & 117 & \%2.60 & 2.97 & 113 & \%2.52 \\
			& Gaussian & 1.36 & 68 & \%2.77 & 2.95 & 159 & \%3.53 & 2.60 & 141 & \%3.15 \\ \hline
			\multirow{2}{*}{0.98} & q-Gaussian & 1.80 & 34 & \%1.38 & 3.79 & 90 & \%1.20 & 3.31 & 88 & \%1.97 \\
			& Gaussian & 1.50 & 56 & \%2.28 & 3.22 & 129 & \%2.87 & 2.84 & 125 & \%2.79 \\ \hline
		\end{tabular}
	\end{center}
\end{table}

\section{Findings and Results}
In this section, we have scrutinized the non-extensive approach to estimate value-at-risk. Also, differences between non-extensive and normal value-at-risk in mature and emerging financial markets over time and various market conditions has been studied.\par

\begin{figure}
	\centering
	\begin{tabular}{@{}c@{}}
		\includegraphics[width=0.3\linewidth]{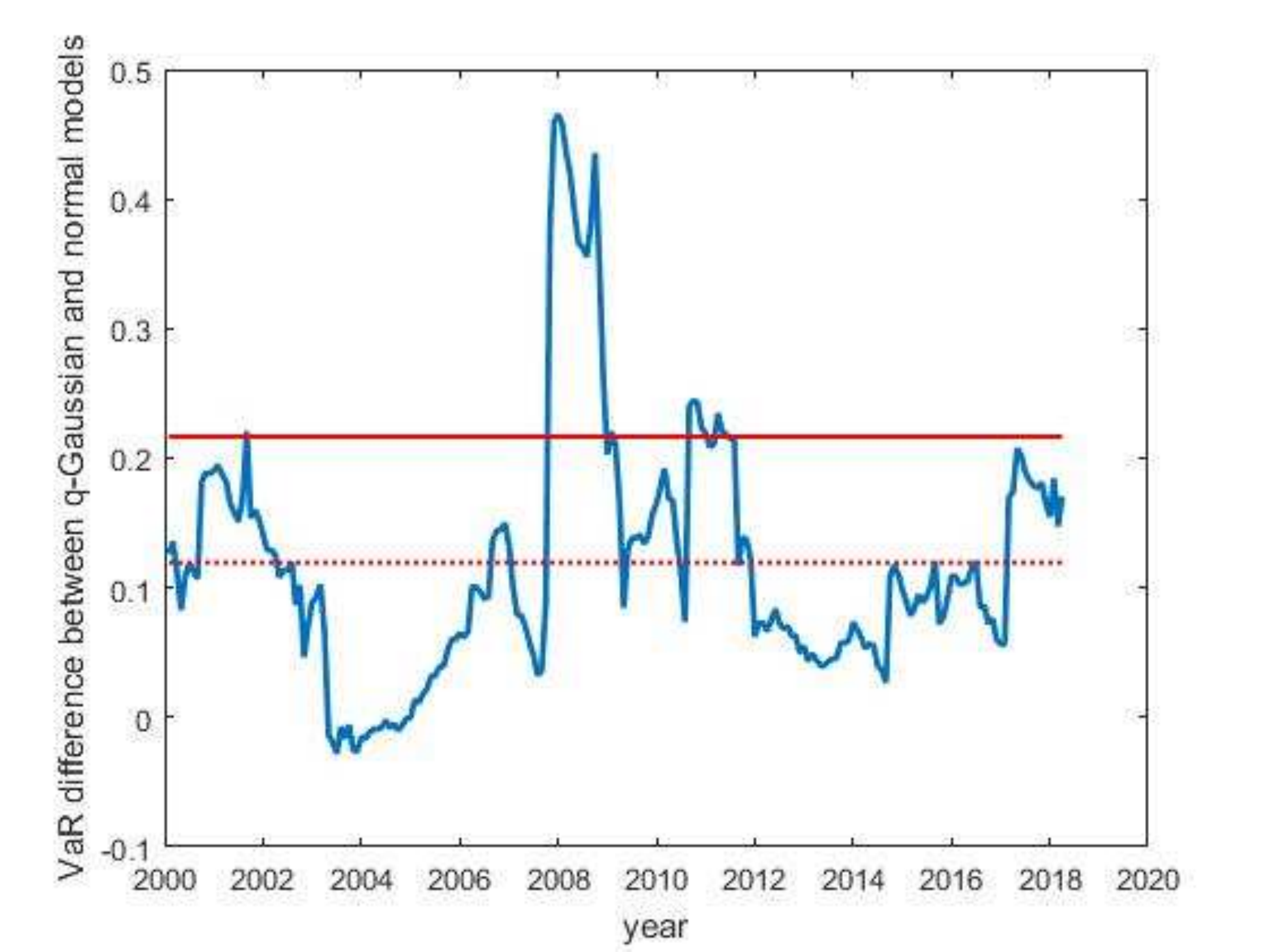}\\[\abovecaptionskip]
		\small DJIA
	\end{tabular}
	\begin{tabular}{@{}c@{}}
		\includegraphics[width=0.3\linewidth]{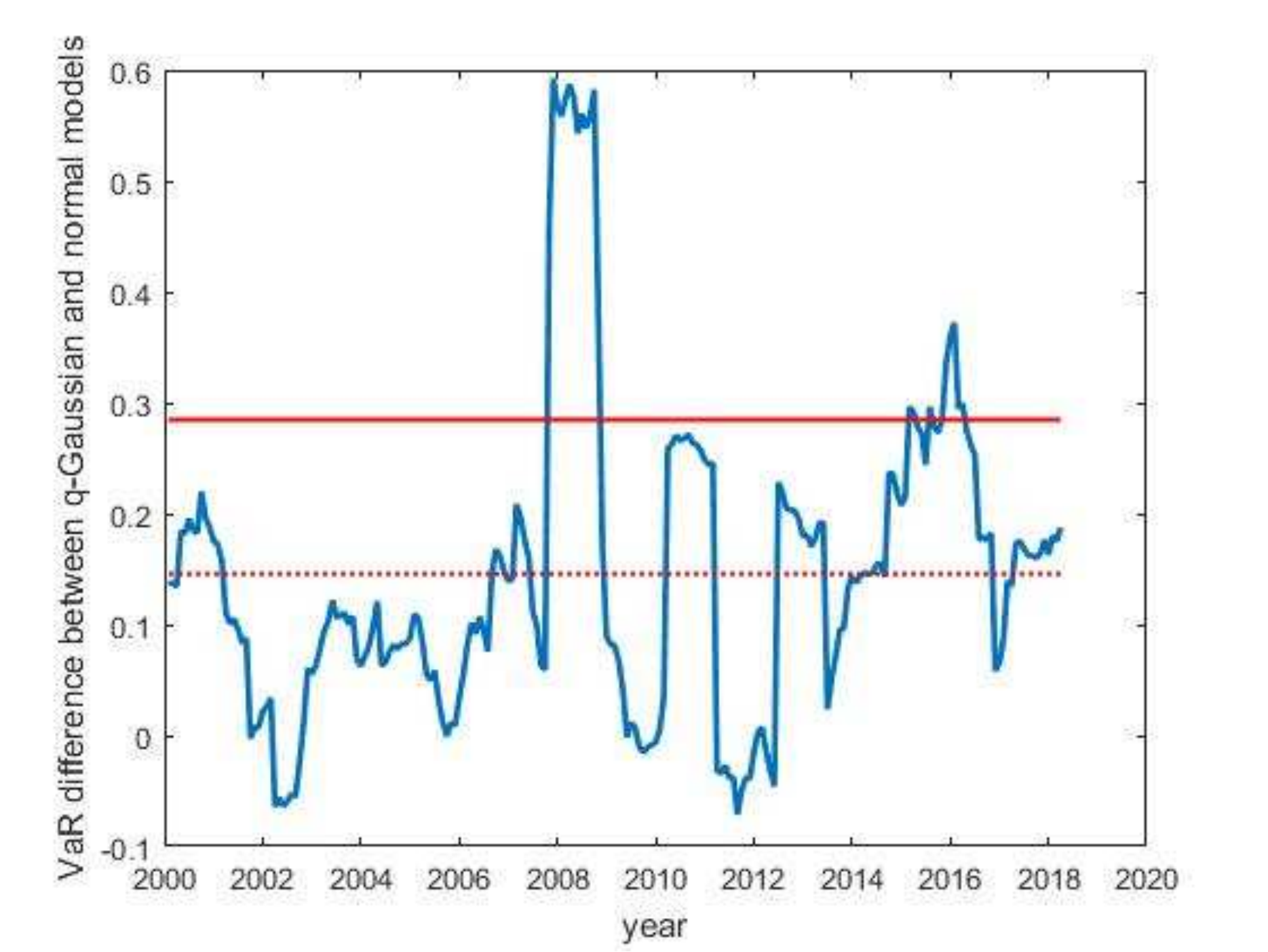}\\[\abovecaptionskip]
		\small NIKKEI 225
	\end{tabular}
	\begin{tabular}{@{}c@{}}
		\includegraphics[width=0.3\linewidth]{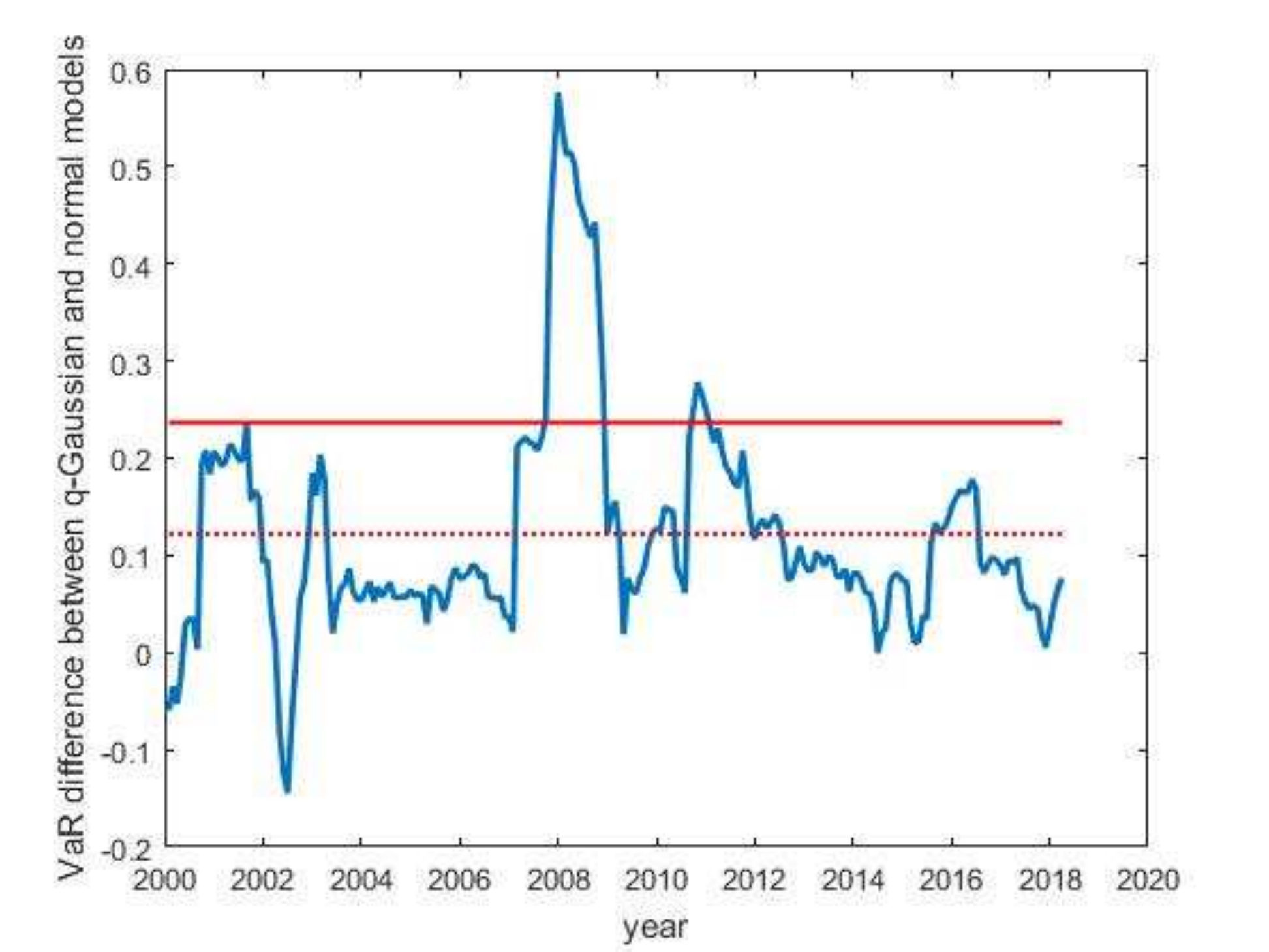}\\[\abovecaptionskip]
		\small DAX
	\end{tabular}
	\begin{tabular}{@{}c@{}}
		\includegraphics[width=0.3\linewidth]{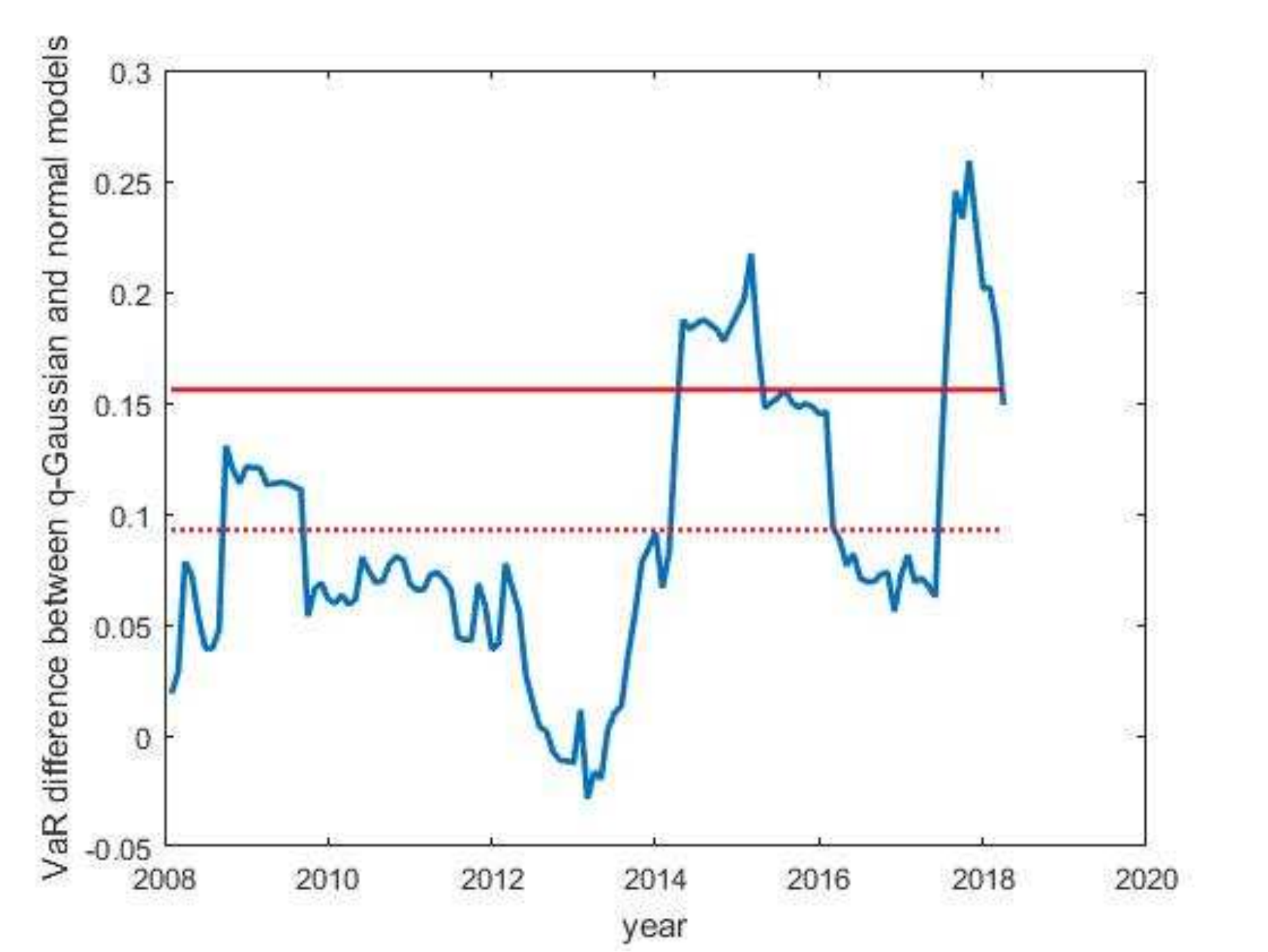}\\[\abovecaptionskip]
		\small TSE
	\end{tabular}
	\begin{tabular}{@{}c@{}}
		\includegraphics[width=0.3\linewidth]{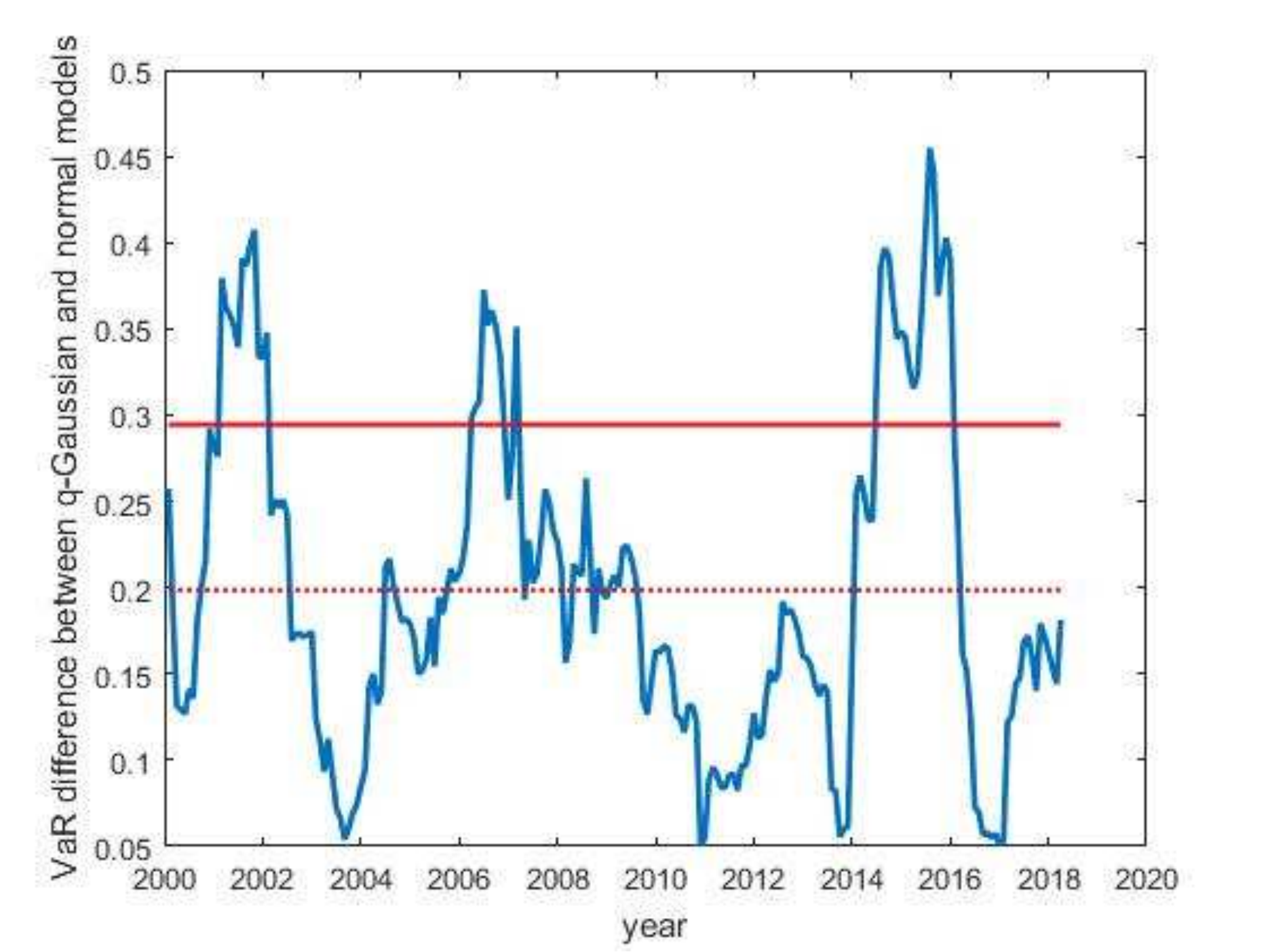}\\[\abovecaptionskip]
		\small SSE
	\end{tabular}
	\begin{tabular}{@{}c@{}}
		\includegraphics[width=0.3\linewidth]{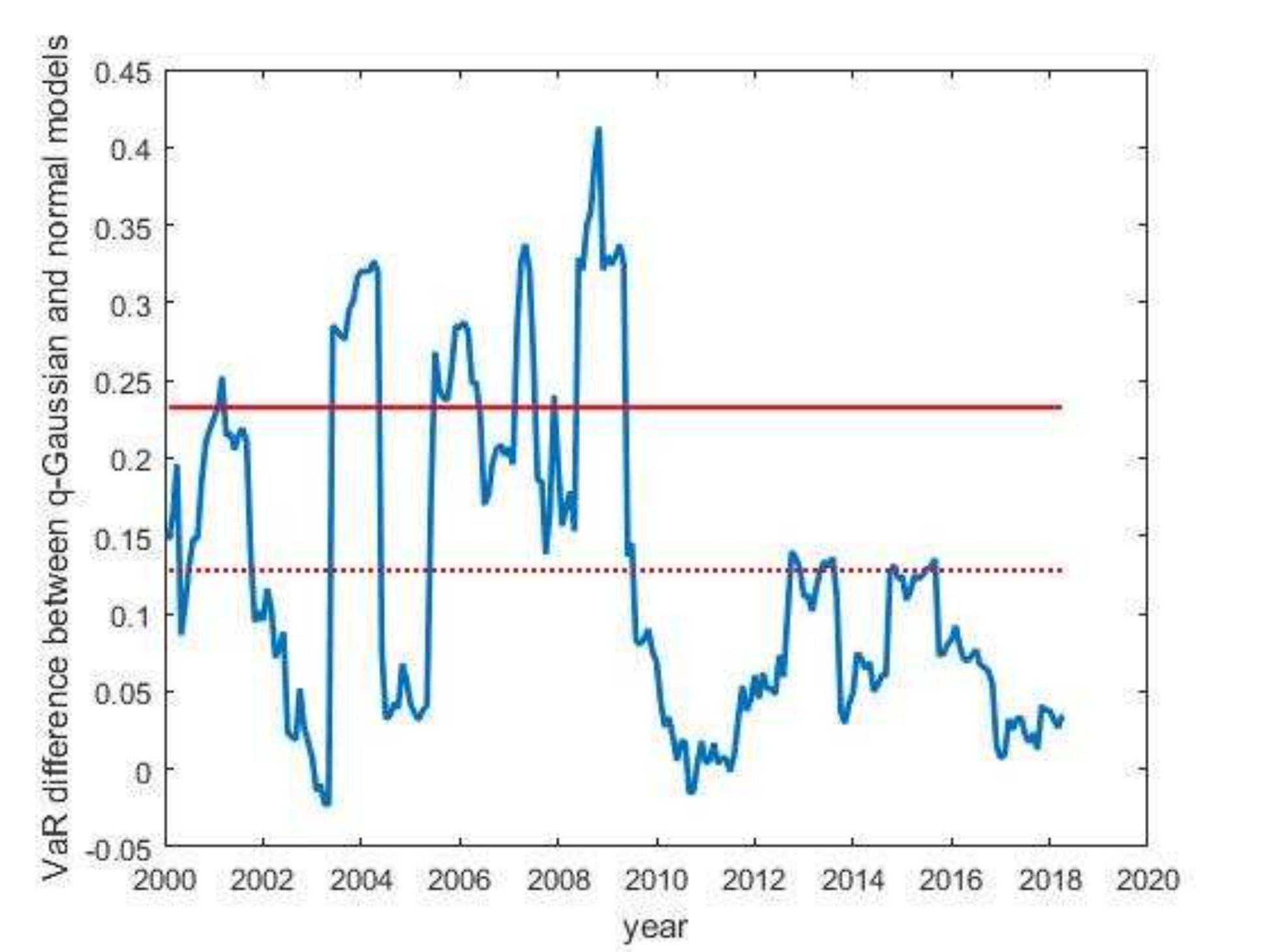}\\[\abovecaptionskip]
		\small BSE
	\end{tabular}
	\caption{\small Difference between q-Gaussian based VaR and normal VaR in mature and emerging markets over time period of study with mean line (dashed) and one standard deviation distance line (continuous)}
\end{figure}

Table 1 and Table 2 represent estimated q parameter, normal and non-extensive value-at-risk and related back-tests in each mature and emerging market. The q parameter estimation approach is based on maximum likelihood estimation (MLE) method. Also, asymptotic error interval (95 percent confidence) of estimated q parameters calculated using the Fisher information of the return time-series. The value-at-risk and back-test results are provided in various confidence levels from 0.95 to 0.98. Here, back-test value is the proportion of cases which violate the estimated value-at-risk to the total number of elements in observed return series. As an example, value-at-risks obtained for DAX with confidence level of 0.97 are 2.76 and 3.14 percent based on normal and q-Gaussian models respectively. In the case of normal condition, approximately 3.43 percent of losses exceed the calculated value-at-risk, which shows underestimation comparing with 3 percent tolerance level. This underestimation is disappeared using q-Gaussian model. In the case of q-Gaussian value-at-risk, almost 2.47 percent of losses are worse than the calculated value-at-risk.\par

\begin{figure}
	\centering
	\begin{tabular}{@{}c@{}}
		\includegraphics[width=\linewidth]{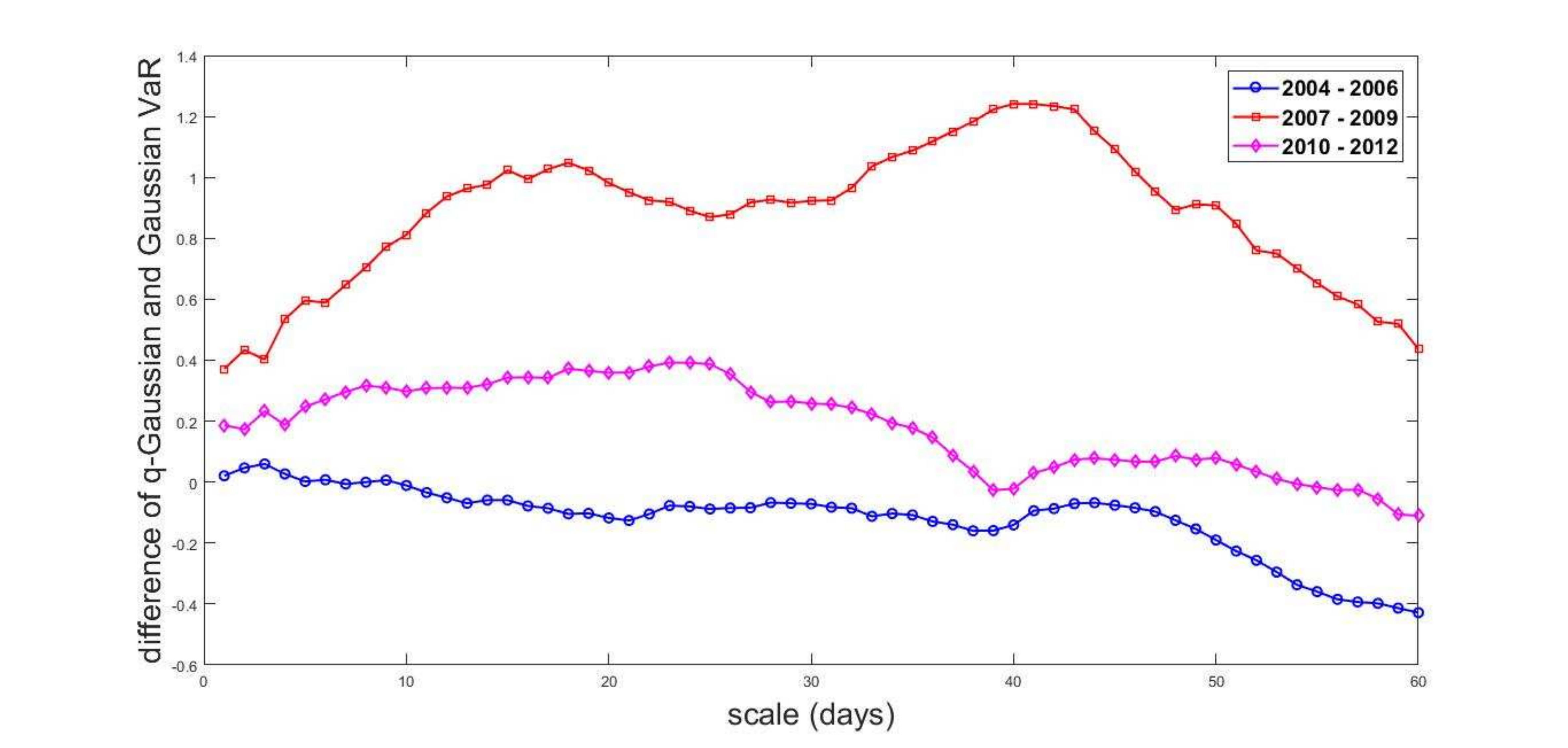}\\[\abovecaptionskip]
		\small DJIA
	\end{tabular}
	\begin{tabular}{@{}c@{}}
		\includegraphics[width=\linewidth]{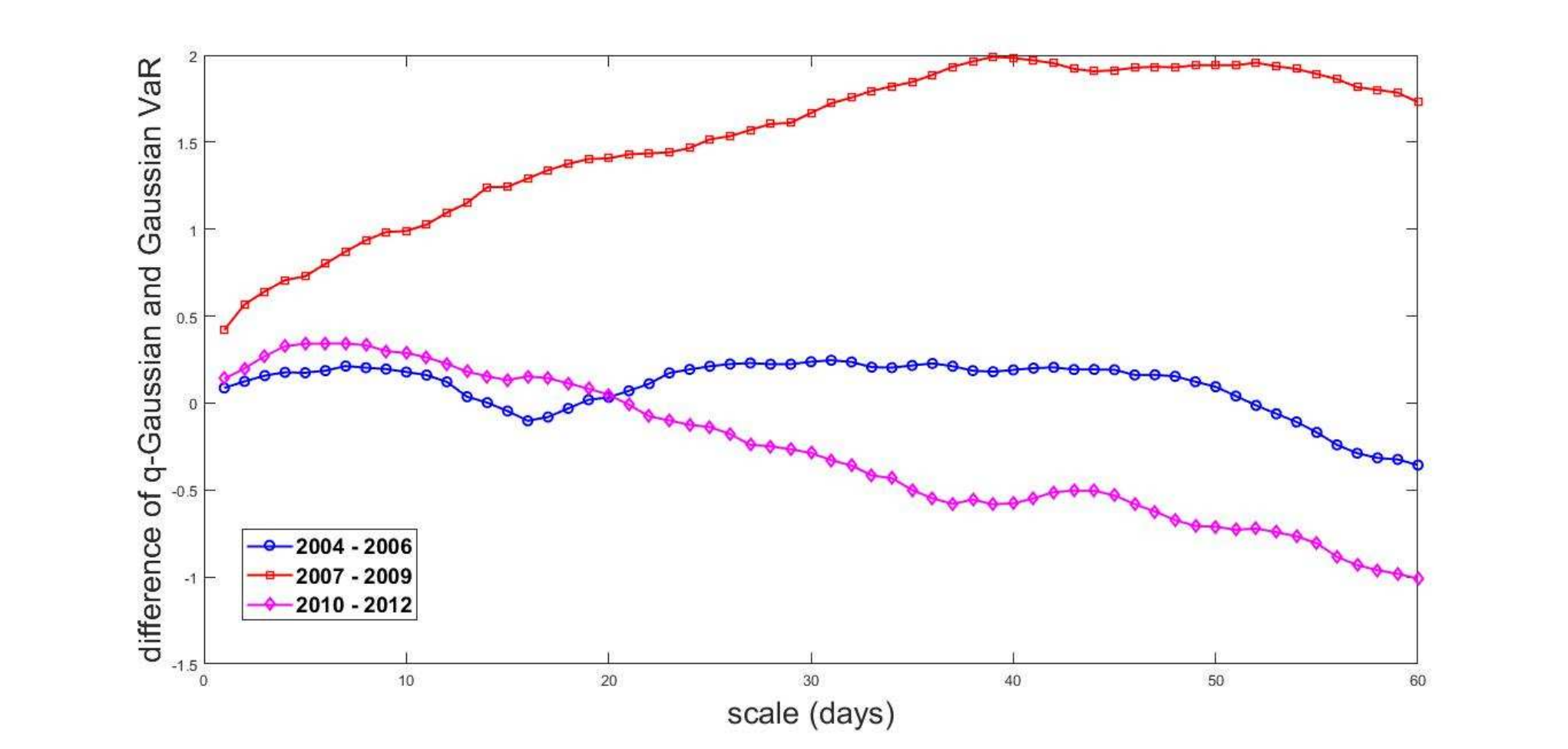}\\[\abovecaptionskip]
		\small NIKKEI 225
	\end{tabular}
	\caption{\small Difference between q-Gaussian based VaR and normal VaR in mature markets with different time scales}
\end{figure}

In Figure 2, the first row shows the difference between non-extensive and normal value-at-risk in three mature markets over time. It is obvious that in the studied mature markets, during the financial crisis of 2007-2009, difference between entropic and normal value-at-risk increases significantly and crosses the one standard deviation line. In this case the differential value can be seen as an indicator of crash in financial market. In addition, all of the studied mature markets follow similar patterns over time. This similarity indicates that, these markets are strongly coupled with each other and act as components of a big economic environment. On the other side, in the studied emerging markets, the difference of non-extensive and normal value-at-risk has random patterns during time. As it is shown in the second row of Figure 2, there is not any clear response during global financial crisis in these markets. Each emerging market has its own unique pattern of differential value. In contrast to the mature markets, each emerging market is a separate ecosystem which interacts with the other markets. This behavior is originated from the domestic economical policies of these countries and their narrow relations with global markets.\par

Figure 3 displays difference of value-at-risk between non-extensive and normal models for DJIA and Nikkei 225 indices, in different time scales. We have provided return series in different time scales from 1 to 60 days, in order to compare short-term and long-term features of financial time-series. The results have shown that during crisis period (2007-2009), as time scale grows larger, difference between non-extensive and normal value-at-risk increases; which is in obvious contrast with pre-crisis (2004-2006) and post-crisis (2010-2012) time periods. This observation exhibits increasing non-Gaussian properties of financial time-series during times of crisis.

\section{Conclusion}
In this research, non-extensive statistical framework has been used to estimate the value-at-risk in mature and emerging markets. We have shown that the q-Gaussian approach is a suitable technique to capture the leptokurtic properties of financial time series. The result of back-tests in both mature and emerging markets showed that non-extensive model of value-at-risk improves the estimated value-at-risk and removes the underestimation happens in normal approach.\par
Also, we have seen that in the mature markets, the difference of value-at-risk between q-Gaussian approach and normal method increases more than one standard deviation during times of crisis. This differential can be used as prognostication for financial crisis in mature markets. The results in the emerging markets are different and the responses during times of financial crisis are not clear. It is observable that all mature markets have same pattern over the time horizon, which shows these markets strongly coupled with each other and are parts of a bigger economic ecosystem. In case of emerging markets, each market has a different random pattern, which indicates they have different environments with different structures.\par
Finally, observing long term features of financial time-series of mature markets revealed that non-Gaussian properties of financial market data, such as extreme values, increase during times of crisis. This observation is compatible with the results of back-tests which has been discussed earlier in this paper.

\bibliography{references}

\begin{thebibliography}{}

\bibitem [\protect \citeauthoryear {%
Acerbi%
\ \BBA {} Tasche%
}{%
Acerbi%
\ \BBA {} Tasche%
}{%
{\protect \APACyear {2002}}%
}]{%
ARTICLE:4}
\APACinsertmetastar {%
ARTICLE:4}%
\begin{APACrefauthors}%
Acerbi, C.%
\BCBT {}\ \BBA {} Tasche, D.%
\end{APACrefauthors}%
\unskip\
\newblock
\APACrefYearMonthDay{2002}{May}{}.
\newblock
{\BBOQ}\APACrefatitle {Expected Shortfall: A Natural Coherent Alternative to
  Value at Risk} {Expected shortfall: A natural coherent alternative to value
  at risk}.{\BBCQ}
\newblock
\APACjournalVolNumPages{Economic Notes}{31}{2}{379-388}.
\PrintBackRefs{\CurrentBib}

\bibitem [\protect \citeauthoryear {%
Basak%
\ \BBA {} Shapiro%
}{%
Basak%
\ \BBA {} Shapiro%
}{%
{\protect \APACyear {2001}}%
}]{%
ARTICLE:5}
\APACinsertmetastar {%
ARTICLE:5}%
\begin{APACrefauthors}%
Basak, S.%
\BCBT {}\ \BBA {} Shapiro, A.%
\end{APACrefauthors}%
\unskip\
\newblock
\APACrefYearMonthDay{2001}{Apr}{}.
\newblock
{\BBOQ}\APACrefatitle {Value-at-Risk-Based Risk Management: Optimal Policies
  and Asset Prices} {Value-at-risk-based risk management: Optimal policies and
  asset prices}.{\BBCQ}
\newblock
\APACjournalVolNumPages{The Review of Financial Studies}{14}{2}{371-405}.
\PrintBackRefs{\CurrentBib}

\bibitem [\protect \citeauthoryear {%
Bil%
, Grech%
\BCBL {}\ \BBA {} Podhajska%
}{%
Bil%
\ \protect \BOthers {.}}{%
{\protect \APACyear {2016}}%
}]{%
ARTICLE:10}
\APACinsertmetastar {%
ARTICLE:10}%
\begin{APACrefauthors}%
Bil, L.%
, Grech, D.%
\BCBL {}\ \BBA {} Podhajska, E.%
\end{APACrefauthors}%
\unskip\
\newblock
\APACrefYearMonthDay{2016}{May}{}.
\newblock
{\BBOQ}\APACrefatitle {Methods of Non-Extensive Statistical Physics in Analysis
  of Price Returns on Polish Stock Market} {Methods of non-extensive
  statistical physics in analysis of price returns on polish stock
  market}.{\BBCQ}
\newblock
\APACjournalVolNumPages{Acta Physica Polonica A}{129}{5}{986-992}.
\PrintBackRefs{\CurrentBib}

\bibitem [\protect \citeauthoryear {%
Borland%
}{%
Borland%
}{%
{\protect \APACyear {2002}}%
}]{%
ARTICLE:11}
\APACinsertmetastar {%
ARTICLE:11}%
\begin{APACrefauthors}%
Borland, L.%
\end{APACrefauthors}%
\unskip\
\newblock
\APACrefYearMonthDay{2002}{Nov}{}.
\newblock
{\BBOQ}\APACrefatitle {A Theory of Non-Gaussian Option Pricing} {A theory of
  non-gaussian option pricing}.{\BBCQ}
\newblock
\APACjournalVolNumPages{Quantitative Finance}{2}{6}{415-431}.
\PrintBackRefs{\CurrentBib}

\bibitem [\protect \citeauthoryear {%
Borland%
\ \BBA {} Bouchaud%
}{%
Borland%
\ \BBA {} Bouchaud%
}{%
{\protect \APACyear {2004}}%
}]{%
ARTICLE:28}
\APACinsertmetastar {%
ARTICLE:28}%
\begin{APACrefauthors}%
Borland, L.%
\BCBT {}\ \BBA {} Bouchaud, J\BHBI P.%
\end{APACrefauthors}%
\unskip\
\newblock
\APACrefYearMonthDay{2004}{Feb}{}.
\newblock
{\BBOQ}\APACrefatitle {A Non-Gaussian Option Pricing Model with Skew} {A
  non-gaussian option pricing model with skew}.{\BBCQ}
\newblock
\APACjournalVolNumPages{Quantitative Finance}{4}{5}{499-514}.
\PrintBackRefs{\CurrentBib}

\bibitem [\protect \citeauthoryear {%
D’Arcangelis%
\ \BBA {} Rotundo%
}{%
D’Arcangelis%
\ \BBA {} Rotundo%
}{%
{\protect \APACyear {2016}}%
}]{%
ARTICLE:32}
\APACinsertmetastar {%
ARTICLE:32}%
\begin{APACrefauthors}%
D’Arcangelis, A\BPBI M.%
\BCBT {}\ \BBA {} Rotundo, G.%
\end{APACrefauthors}%
\unskip\
\newblock
\APACrefYearMonthDay{2016}{Sep}{}.
\newblock
{\BBOQ}\APACrefatitle {Complex Networks in Finance} {Complex networks in
  finance}.{\BBCQ}
\newblock
\APACjournalVolNumPages{Complex Networks and Dynamics. Lecture Notes in
  Economics and Mathematical Systems, Springer, Cham}{683}{}{209-235}.
\PrintBackRefs{\CurrentBib}

\bibitem [\protect \citeauthoryear {%
Fama%
}{%
Fama%
}{%
{\protect \APACyear {1965}}%
}]{%
ARTICLE:15}
\APACinsertmetastar {%
ARTICLE:15}%
\begin{APACrefauthors}%
Fama, E\BPBI F.%
\end{APACrefauthors}%
\unskip\
\newblock
\APACrefYearMonthDay{1965}{Jan}{}.
\newblock
{\BBOQ}\APACrefatitle {The Behavior of Stock-Market Prices} {The behavior of
  stock-market prices}.{\BBCQ}
\newblock
\APACjournalVolNumPages{The Journal of Business}{38}{1}{34-105}.
\PrintBackRefs{\CurrentBib}

\bibitem [\protect \citeauthoryear {%
Farmer%
\ \protect \BOthers {.}}{%
Farmer%
\ \protect \BOthers {.}}{%
{\protect \APACyear {2012}}%
}]{%
ARTICLE:31}
\APACinsertmetastar {%
ARTICLE:31}%
\begin{APACrefauthors}%
Farmer, J\BPBI D.%
, Gallegati, M.%
, Hommes, C.%
, Kirman, A.%
, Ormerod, P.%
, Cincotti, S.%
\BDBL {}Helbing, D.%
\end{APACrefauthors}%
\unskip\
\newblock
\APACrefYearMonthDay{2012}{Dec}{}.
\newblock
{\BBOQ}\APACrefatitle {A Complex Systems Approach to Constructing Better Models
  for Managing Financial Markets and the Economy} {A complex systems approach
  to constructing better models for managing financial markets and the
  economy}.{\BBCQ}
\newblock
\APACjournalVolNumPages{The European Physical Journal Special
  Topics}{214}{}{295–324}.
\PrintBackRefs{\CurrentBib}

\bibitem [\protect \citeauthoryear {%
Furuichi%
}{%
Furuichi%
}{%
{\protect \APACyear {2009}}%
}]{%
ARTICLE:17}
\APACinsertmetastar {%
ARTICLE:17}%
\begin{APACrefauthors}%
Furuichi, S.%
\end{APACrefauthors}%
\unskip\
\newblock
\APACrefYearMonthDay{2009}{Jan}{}.
\newblock
{\BBOQ}\APACrefatitle {On the Maximum Entropy Principle and the Minimization of
  the Fisher Information in Tsallis Statistics} {On the maximum entropy
  principle and the minimization of the fisher information in tsallis
  statistics}.{\BBCQ}
\newblock
\APACjournalVolNumPages{Journal of Mathematical Physics}{50}{1}{013303}.
\PrintBackRefs{\CurrentBib}

\bibitem [\protect \citeauthoryear {%
{Gençay}%
\ \BBA {} Gradojevic%
}{%
{Gençay}%
\ \BBA {} Gradojevic%
}{%
{\protect \APACyear {2017}}%
}]{%
ARTICLE:9}
\APACinsertmetastar {%
ARTICLE:9}%
\begin{APACrefauthors}%
{Gençay}, R.%
\BCBT {}\ \BBA {} Gradojevic, N.%
\end{APACrefauthors}%
\unskip\
\newblock
\APACrefYearMonthDay{2017}{May}{}.
\newblock
{\BBOQ}\APACrefatitle {The Tale of Two Financial Crises: An Entropic
  Perspective} {The tale of two financial crises: An entropic
  perspective}.{\BBCQ}
\newblock
\APACjournalVolNumPages{Entropy}{19}{6}{244}.
\PrintBackRefs{\CurrentBib}

\bibitem [\protect \citeauthoryear {%
Gradojevic%
\ \BBA {} {Gençay}%
}{%
Gradojevic%
\ \BBA {} {Gençay}%
}{%
{\protect \APACyear {2008}}%
}]{%
ARTICLE:22}
\APACinsertmetastar {%
ARTICLE:22}%
\begin{APACrefauthors}%
Gradojevic, N.%
\BCBT {}\ \BBA {} {Gençay}, R.%
\end{APACrefauthors}%
\unskip\
\newblock
\APACrefYearMonthDay{2008}{}{}.
\newblock
{\BBOQ}\APACrefatitle {Overnight Interest Rates and Aggregate Market
  Expectations} {Overnight interest rates and aggregate market
  expectations}.{\BBCQ}
\newblock
\APACjournalVolNumPages{Economics Letters}{100}{}{27-30}.
\PrintBackRefs{\CurrentBib}

\bibitem [\protect \citeauthoryear {%
Gradojevic%
\ \BBA {} {Gençay}%
}{%
Gradojevic%
\ \BBA {} {Gençay}%
}{%
{\protect \APACyear {2011}}%
}]{%
ARTICLE:21}
\APACinsertmetastar {%
ARTICLE:21}%
\begin{APACrefauthors}%
Gradojevic, N.%
\BCBT {}\ \BBA {} {Gençay}, R.%
\end{APACrefauthors}%
\unskip\
\newblock
\APACrefYearMonthDay{2011}{Oct}{}.
\newblock
{\BBOQ}\APACrefatitle {Financial Applications of Nonextensive Entropy}
  {Financial applications of nonextensive entropy}.{\BBCQ}
\newblock
\APACjournalVolNumPages{IEEE Signal Processing Magazine}{28}{5}{116-141}.
\PrintBackRefs{\CurrentBib}

\bibitem [\protect \citeauthoryear {%
Hogan%
\ \BBA {} Warren%
}{%
Hogan%
\ \BBA {} Warren%
}{%
{\protect \APACyear {1974}}%
}]{%
ARTICLE:2}
\APACinsertmetastar {%
ARTICLE:2}%
\begin{APACrefauthors}%
Hogan, W\BPBI W.%
\BCBT {}\ \BBA {} Warren, J\BPBI M.%
\end{APACrefauthors}%
\unskip\
\newblock
\APACrefYearMonthDay{1974}{Jan}{}.
\newblock
{\BBOQ}\APACrefatitle {Toward the Development of an Equilibrium Capital-Market
  Model Based on Semivariance} {Toward the development of an equilibrium
  capital-market model based on semivariance}.{\BBCQ}
\newblock
\APACjournalVolNumPages{The Journal of Financial and Quantitative
  Analysis}{9}{1}{1-11}.
\PrintBackRefs{\CurrentBib}

\bibitem [\protect \citeauthoryear {%
Hosseini%
, Iranmanesh%
, Javaran%
\BCBL {}\ \BBA {} Zadehgol%
}{%
Hosseini%
\ \protect \BOthers {.}}{%
{\protect \APACyear {2017}}%
}]{%
ARTICLE:34}
\APACinsertmetastar {%
ARTICLE:34}%
\begin{APACrefauthors}%
Hosseini, A.%
, Iranmanesh, M.%
, Javaran, E\BPBI J.%
\BCBL {}\ \BBA {} Zadehgol, A.%
\end{APACrefauthors}%
\unskip\
\newblock
\APACrefYearMonthDay{2017}{}{}.
\newblock
{\BBOQ}\APACrefatitle {Application of Lattice Kinetic Models with Tsallis
  Entropy in Simulating Fluid Flow Through Porous Media} {Application of
  lattice kinetic models with tsallis entropy in simulating fluid flow through
  porous media}.{\BBCQ}
\newblock
\APACjournalVolNumPages{International Journal of Modern Physics C}{28}{09}{}.
\PrintBackRefs{\CurrentBib}

\bibitem [\protect \citeauthoryear {%
Hosseiny%
, Bahrami%
, Palestrini%
\BCBL {}\ \BBA {} Gallegati%
}{%
Hosseiny%
\ \protect \BOthers {.}}{%
{\protect \APACyear {2016}}%
}]{%
10.1371/journal.pone.0160363}
\APACinsertmetastar {%
10.1371/journal.pone.0160363}%
\begin{APACrefauthors}%
Hosseiny, A.%
, Bahrami, M.%
, Palestrini, A.%
\BCBL {}\ \BBA {} Gallegati, M.%
\end{APACrefauthors}%
\unskip\
\newblock
\APACrefYearMonthDay{2016}{10}{}.
\newblock
{\BBOQ}\APACrefatitle {Metastable Features of Economic Networks and Responses
  to Exogenous Shocks} {Metastable features of economic networks and responses
  to exogenous shocks}.{\BBCQ}
\newblock
\APACjournalVolNumPages{PLOS ONE}{11}{10}{1-22}.
\newblock
\begin{APACrefURL} \url{https://doi.org/10.1371/journal.pone.0160363}
  \end{APACrefURL}
\newblock
\begin{APACrefDOI} \doi{10.1371/journal.pone.0160363} \end{APACrefDOI}
\PrintBackRefs{\CurrentBib}

\bibitem [\protect \citeauthoryear {%
Jorion%
}{%
Jorion%
}{%
{\protect \APACyear {2006}}%
}]{%
BOOK:1}
\APACinsertmetastar {%
BOOK:1}%
\begin{APACrefauthors}%
Jorion, P.%
\end{APACrefauthors}%
\unskip\
\newblock
\APACrefYear{2006}.
\newblock
\APACrefbtitle {Value at Risk: The New Benchmark for Managing Financial Risk}
  {Value at risk: The new benchmark for managing financial risk}\
  (\PrintOrdinal{third}\ \BEd).
\newblock
\APACaddressPublisher{}{McGraw-Hill}.
\PrintBackRefs{\CurrentBib}

\bibitem [\protect \citeauthoryear {%
Kittiakarasakun%
\ \BBA {} Tse%
}{%
Kittiakarasakun%
\ \BBA {} Tse%
}{%
{\protect \APACyear {2011}}%
}]{%
ARTICLE:7}
\APACinsertmetastar {%
ARTICLE:7}%
\begin{APACrefauthors}%
Kittiakarasakun, J.%
\BCBT {}\ \BBA {} Tse, Y.%
\end{APACrefauthors}%
\unskip\
\newblock
\APACrefYearMonthDay{2011}{Jun}{}.
\newblock
{\BBOQ}\APACrefatitle {Modeling the Fat Tails in Asian Stock Markets} {Modeling
  the fat tails in asian stock markets}.{\BBCQ}
\newblock
\APACjournalVolNumPages{International Review of Economics \&
  Finance}{20}{3}{430-440}.
\PrintBackRefs{\CurrentBib}

\bibitem [\protect \citeauthoryear {%
Kozaki%
\ \BBA {} Sato%
}{%
Kozaki%
\ \BBA {} Sato%
}{%
{\protect \APACyear {2008}}%
}]{%
ARTICLE:12}
\APACinsertmetastar {%
ARTICLE:12}%
\begin{APACrefauthors}%
Kozaki, M.%
\BCBT {}\ \BBA {} Sato, A\BHBI H.%
\end{APACrefauthors}%
\unskip\
\newblock
\APACrefYearMonthDay{2008}{Feb}{}.
\newblock
{\BBOQ}\APACrefatitle {Applications of the Beck Model to Stock Markets:
  Value-at-Risk and Portfolio Risk Assessment} {Applications of the beck model
  to stock markets: Value-at-risk and portfolio risk assessment}.{\BBCQ}
\newblock
\APACjournalVolNumPages{Physica A: Statistical Mechanics and its
  Applications}{387}{5-6}{1225-1246}.
\PrintBackRefs{\CurrentBib}

\bibitem [\protect \citeauthoryear {%
Lux%
}{%
Lux%
}{%
{\protect \APACyear {1998}}%
}]{%
ARTICLE:6}
\APACinsertmetastar {%
ARTICLE:6}%
\begin{APACrefauthors}%
Lux, T.%
\end{APACrefauthors}%
\unskip\
\newblock
\APACrefYearMonthDay{1998}{Jan}{}.
\newblock
{\BBOQ}\APACrefatitle {The Socio-Economic Dynamics of Speculative Markets:
  Interacting Agents, Chaos, and the Fat Tails of Return Distributions} {The
  socio-economic dynamics of speculative markets: Interacting agents, chaos,
  and the fat tails of return distributions}.{\BBCQ}
\newblock
\APACjournalVolNumPages{Journal of Economic Behavior \&
  Organization}{33}{2}{143-165}.
\PrintBackRefs{\CurrentBib}

\bibitem [\protect \citeauthoryear {%
Markowitz%
}{%
Markowitz%
}{%
{\protect \APACyear {1952}}%
}]{%
ARTICLE:1}
\APACinsertmetastar {%
ARTICLE:1}%
\begin{APACrefauthors}%
Markowitz, H.%
\end{APACrefauthors}%
\unskip\
\newblock
\APACrefYearMonthDay{1952}{Mar}{}.
\newblock
{\BBOQ}\APACrefatitle {Portfolio Selection} {Portfolio selection}.{\BBCQ}
\newblock
\APACjournalVolNumPages{The Journal of Finance}{7}{1}{77-91}.
\PrintBackRefs{\CurrentBib}

\bibitem [\protect \citeauthoryear {%
Namaki%
, Lai%
, Jafari%
, Raei%
\BCBL {}\ \BBA {} Tehrani%
}{%
Namaki%
\ \protect \BOthers {.}}{%
{\protect \APACyear {2013}}%
}]{%
ARTICLE:14}
\APACinsertmetastar {%
ARTICLE:14}%
\begin{APACrefauthors}%
Namaki, A.%
, Lai, Z\BPBI K.%
, Jafari, G\BPBI R.%
, Raei, R.%
\BCBL {}\ \BBA {} Tehrani, R.%
\end{APACrefauthors}%
\unskip\
\newblock
\APACrefYearMonthDay{2013}{Jul}{}.
\newblock
{\BBOQ}\APACrefatitle {Comparing Emerging and Mature Markets During Times of
  Crisis: A Non-Extensive Statistical Approach} {Comparing emerging and mature
  markets during times of crisis: A non-extensive statistical approach}.{\BBCQ}
\newblock
\APACjournalVolNumPages{Physica A: Statistical Mechanics and its
  Applications}{392}{14}{3039-3044}.
\PrintBackRefs{\CurrentBib}

\bibitem [\protect \citeauthoryear {%
Namaki%
, Raei%
, Asadi%
\BCBL {}\ \BBA {} Hajihasani%
}{%
Namaki%
\ \protect \BOthers {.}}{%
{\protect \APACyear {2019}}%
}]{%
ARTICLE:30}
\APACinsertmetastar {%
ARTICLE:30}%
\begin{APACrefauthors}%
Namaki, A.%
, Raei, R.%
, Asadi, N.%
\BCBL {}\ \BBA {} Hajihasani, A.%
\end{APACrefauthors}%
\unskip\
\newblock
\APACrefYearMonthDay{2019}{}{}.
\newblock
{\BBOQ}\APACrefatitle {Analysis of Iran Banking Sector by Multi-Layer Approach}
  {Analysis of iran banking sector by multi-layer approach}.{\BBCQ}
\newblock
\APACjournalVolNumPages{Iranian Journal of Finance}{3}{1}{73-89}.
\PrintBackRefs{\CurrentBib}

\bibitem [\protect \citeauthoryear {%
Namaki%
, Raei%
\BCBL {}\ \BBA {} Jafari%
}{%
Namaki%
\ \protect \BOthers {.}}{%
{\protect \APACyear {2011}}%
}]{%
ARTICLE:35}
\APACinsertmetastar {%
ARTICLE:35}%
\begin{APACrefauthors}%
Namaki, A.%
, Raei, R.%
\BCBL {}\ \BBA {} Jafari, G\BPBI R.%
\end{APACrefauthors}%
\unskip\
\newblock
\APACrefYearMonthDay{2011}{}{}.
\newblock
{\BBOQ}\APACrefatitle {Comparing Tehran Stock Exchange as an Emerging Market
  with a Mature Market by Random Matrix Approach} {Comparing tehran stock
  exchange as an emerging market with a mature market by random matrix
  approach}.{\BBCQ}
\newblock
\APACjournalVolNumPages{International Journal of Modern Physics
  C}{22}{04}{371-383}.
\PrintBackRefs{\CurrentBib}

\bibitem [\protect \citeauthoryear {%
Nawrocki%
}{%
Nawrocki%
}{%
{\protect \APACyear {1992}}%
}]{%
ARTICLE:3}
\APACinsertmetastar {%
ARTICLE:3}%
\begin{APACrefauthors}%
Nawrocki, D\BPBI N.%
\end{APACrefauthors}%
\unskip\
\newblock
\APACrefYearMonthDay{1992}{}{}.
\newblock
{\BBOQ}\APACrefatitle {The Characteristics of portfolios selected by n-degree
  Lower Partial Moment} {The characteristics of portfolios selected by n-degree
  lower partial moment}.{\BBCQ}
\newblock
\APACjournalVolNumPages{International Review of Financial
  Analysis}{1}{3}{195-209}.
\PrintBackRefs{\CurrentBib}

\bibitem [\protect \citeauthoryear {%
NicholasTaleb%
}{%
NicholasTaleb%
}{%
{\protect \APACyear {2009}}%
}]{%
ARTICLE:16}
\APACinsertmetastar {%
ARTICLE:16}%
\begin{APACrefauthors}%
NicholasTaleb, N.%
\end{APACrefauthors}%
\unskip\
\newblock
\APACrefYearMonthDay{2009}{Oct}{}.
\newblock
{\BBOQ}\APACrefatitle {Errors, Robustness, and the Fourth Quadrant} {Errors,
  robustness, and the fourth quadrant}.{\BBCQ}
\newblock
\APACjournalVolNumPages{International Journal of Forecasting}{25}{4}{744-759}.
\PrintBackRefs{\CurrentBib}

\bibitem [\protect \citeauthoryear {%
Queirós%
}{%
Queirós%
}{%
{\protect \APACyear {2005}}%
}]{%
ARTICLE:26}
\APACinsertmetastar {%
ARTICLE:26}%
\begin{APACrefauthors}%
Queirós, S\BPBI M\BPBI D.%
\end{APACrefauthors}%
\unskip\
\newblock
\APACrefYearMonthDay{2005}{}{}.
\newblock
{\BBOQ}\APACrefatitle {On non-Gaussianity and Dependence in Financial Time
  Series: a Nonextensive Approach} {On non-gaussianity and dependence in
  financial time series: a nonextensive approach}.{\BBCQ}
\newblock
\APACjournalVolNumPages{Quantitative Finance}{5}{5}{475-487}.
\PrintBackRefs{\CurrentBib}

\bibitem [\protect \citeauthoryear {%
Queirós%
\ \BBA {} Tsallis%
}{%
Queirós%
\ \BBA {} Tsallis%
}{%
{\protect \APACyear {2005}}%
}]{%
ARTICLE:29}
\APACinsertmetastar {%
ARTICLE:29}%
\begin{APACrefauthors}%
Queirós, S\BPBI M\BPBI D.%
\BCBT {}\ \BBA {} Tsallis, C.%
\end{APACrefauthors}%
\unskip\
\newblock
\APACrefYearMonthDay{2005}{}{}.
\newblock
{\BBOQ}\APACrefatitle {Bridging a Paradigmatic Financial Model and Nonextensive
  Entropy} {Bridging a paradigmatic financial model and nonextensive
  entropy}.{\BBCQ}
\newblock
\APACjournalVolNumPages{EPL (Europhysics Letters)}{69}{6}{893}.
\PrintBackRefs{\CurrentBib}

\bibitem [\protect \citeauthoryear {%
Raei%
, Namaki%
\BCBL {}\ \BBA {} Vahabi%
}{%
Raei%
\ \protect \BOthers {.}}{%
{\protect \APACyear {2019}}%
}]{%
ARTICLE:36}
\APACinsertmetastar {%
ARTICLE:36}%
\begin{APACrefauthors}%
Raei, R.%
, Namaki, A.%
\BCBL {}\ \BBA {} Vahabi, H.%
\end{APACrefauthors}%
\unskip\
\newblock
\APACrefYearMonthDay{2019}{}{}.
\newblock
{\BBOQ}\APACrefatitle {Analysis of Collective Behavior of Iran Banking Sector
  by Random Matrix Theory} {Analysis of collective behavior of iran banking
  sector by random matrix theory}.{\BBCQ}
\newblock
\APACjournalVolNumPages{Iranian Journal of Finance}{3}{4}{60-75}.
\PrintBackRefs{\CurrentBib}

\bibitem [\protect \citeauthoryear {%
Resti%
\ \BBA {} Sironi%
}{%
Resti%
\ \BBA {} Sironi%
}{%
{\protect \APACyear {2007}}%
}]{%
BOOK:2}
\APACinsertmetastar {%
BOOK:2}%
\begin{APACrefauthors}%
Resti, A.%
\BCBT {}\ \BBA {} Sironi, A.%
\end{APACrefauthors}%
\unskip\
\newblock
\APACrefYear{2007}.
\newblock
\APACrefbtitle {Risk Management and Shareholders' Value in Banking: From Risk
  Measurement Models to Capital Allocation Policies} {Risk management and
  shareholders' value in banking: From risk measurement models to capital
  allocation policies}\ (\PrintOrdinal{first}\ \BEd).
\newblock
\APACaddressPublisher{}{Wiley}.
\PrintBackRefs{\CurrentBib}

\bibitem [\protect \citeauthoryear {%
Schweitzer%
\ \protect \BOthers {.}}{%
Schweitzer%
\ \protect \BOthers {.}}{%
{\protect \APACyear {2009}}%
}]{%
ARTICLE:33}
\APACinsertmetastar {%
ARTICLE:33}%
\begin{APACrefauthors}%
Schweitzer, F.%
, Fagiolo, G.%
, Sornette, D.%
, Vega-Redondo, F.%
, Vespignani, A.%
\BCBL {}\ \BBA {} White, D\BPBI R.%
\end{APACrefauthors}%
\unskip\
\newblock
\APACrefYearMonthDay{2009}{Jul}{}.
\newblock
{\BBOQ}\APACrefatitle {Economic Networks: The New Challenges} {Economic
  networks: The new challenges}.{\BBCQ}
\newblock
\APACjournalVolNumPages{Science}{325}{5939}{422-425}.
\PrintBackRefs{\CurrentBib}

\bibitem [\protect \citeauthoryear {%
Sornette%
}{%
Sornette%
}{%
{\protect \APACyear {2002}}%
}]{%
ARTICLE:24}
\APACinsertmetastar {%
ARTICLE:24}%
\begin{APACrefauthors}%
Sornette, D.%
\end{APACrefauthors}%
\unskip\
\newblock
\APACrefYearMonthDay{2002}{Feb}{}.
\newblock
{\BBOQ}\APACrefatitle {Predictability of Catastrophic Events: Material Rupture,
  Earthquakes, Turbulence, Financial Crashes, and Human Birth} {Predictability
  of catastrophic events: Material rupture, earthquakes, turbulence, financial
  crashes, and human birth}.{\BBCQ}
\newblock
\APACjournalVolNumPages{Proceedings of the National Academy of Sciences of the
  United States of America}{99}{}{2522-2529}.
\PrintBackRefs{\CurrentBib}

\bibitem [\protect \citeauthoryear {%
Sornette%
}{%
Sornette%
}{%
{\protect \APACyear {2009}}%
}]{%
ARTICLE:23}
\APACinsertmetastar {%
ARTICLE:23}%
\begin{APACrefauthors}%
Sornette, D.%
\end{APACrefauthors}%
\unskip\
\newblock
\APACrefYearMonthDay{2009}{}{}.
\newblock
{\BBOQ}\APACrefatitle {Dragon-Kings, Black Swans and the Prediction of Crises}
  {Dragon-kings, black swans and the prediction of crises}.{\BBCQ}
\newblock
\APACjournalVolNumPages{International Journal of Terraspace Science and
  Engineering}{2}{1}{1-18}.
\PrintBackRefs{\CurrentBib}

\bibitem [\protect \citeauthoryear {%
Tsallis%
}{%
Tsallis%
}{%
{\protect \APACyear {1988}}%
}]{%
ARTICLE:20}
\APACinsertmetastar {%
ARTICLE:20}%
\begin{APACrefauthors}%
Tsallis, C.%
\end{APACrefauthors}%
\unskip\
\newblock
\APACrefYearMonthDay{1988}{Jul}{}.
\newblock
{\BBOQ}\APACrefatitle {Possible Generalization of Boltzmann-Gibbs Statistics}
  {Possible generalization of boltzmann-gibbs statistics}.{\BBCQ}
\newblock
\APACjournalVolNumPages{Journal of Statistical Physics}{52}{1-2}{479-487}.
\PrintBackRefs{\CurrentBib}

\bibitem [\protect \citeauthoryear {%
Tsallis%
}{%
Tsallis%
}{%
{\protect \APACyear {1995}}%
}]{%
ARTICLE:18}
\APACinsertmetastar {%
ARTICLE:18}%
\begin{APACrefauthors}%
Tsallis, C.%
\end{APACrefauthors}%
\unskip\
\newblock
\APACrefYearMonthDay{1995}{}{}.
\newblock
{\BBOQ}\APACrefatitle {Some Comments on Boltzmann-Gibbs Statistical Mechanics}
  {Some comments on boltzmann-gibbs statistical mechanics}.{\BBCQ}
\newblock
\APACjournalVolNumPages{Chaos, Solitons \& Fractals}{6}{}{539-559}.
\PrintBackRefs{\CurrentBib}

\bibitem [\protect \citeauthoryear {%
Tsallis%
}{%
Tsallis%
}{%
{\protect \APACyear {1998}}%
}]{%
ARTICLE:8}
\APACinsertmetastar {%
ARTICLE:8}%
\begin{APACrefauthors}%
Tsallis, C.%
\end{APACrefauthors}%
\unskip\
\newblock
\APACrefYearMonthDay{1998}{Jul}{}.
\newblock
{\BBOQ}\APACrefatitle {Possible Generalization of Boltzmann-Gibbs Statistics}
  {Possible generalization of boltzmann-gibbs statistics}.{\BBCQ}
\newblock
\APACjournalVolNumPages{Journal of Statistical Physics}{52}{1-2}{479-487}.
\PrintBackRefs{\CurrentBib}

\bibitem [\protect \citeauthoryear {%
Tsallis%
}{%
Tsallis%
}{%
{\protect \APACyear {2002}}%
}]{%
ARTICLE:19}
\APACinsertmetastar {%
ARTICLE:19}%
\begin{APACrefauthors}%
Tsallis, C.%
\end{APACrefauthors}%
\unskip\
\newblock
\APACrefYearMonthDay{2002}{Mar}{}.
\newblock
{\BBOQ}\APACrefatitle {Entropic Nonextensivity: A Possible Measure of
  Complexity} {Entropic nonextensivity: A possible measure of
  complexity}.{\BBCQ}
\newblock
\APACjournalVolNumPages{Chaos, Solitons \& Fractals}{13}{3}{371-391}.
\PrintBackRefs{\CurrentBib}

\bibitem [\protect \citeauthoryear {%
Tsallis%
}{%
Tsallis%
}{%
{\protect \APACyear {2017}}%
}]{%
ARTICLE:25}
\APACinsertmetastar {%
ARTICLE:25}%
\begin{APACrefauthors}%
Tsallis, C.%
\end{APACrefauthors}%
\unskip\
\newblock
\APACrefYearMonthDay{2017}{Aug}{}.
\newblock
{\BBOQ}\APACrefatitle {Economics and Finance: q-Statistical Stylized Features
  Galore} {Economics and finance: q-statistical stylized features
  galore}.{\BBCQ}
\newblock
\APACjournalVolNumPages{Entropy}{19}{9}{457}.
\PrintBackRefs{\CurrentBib}

\bibitem [\protect \citeauthoryear {%
Vellekoop%
\ \BBA {} Nieuwenhuis%
}{%
Vellekoop%
\ \BBA {} Nieuwenhuis%
}{%
{\protect \APACyear {2007}}%
}]{%
ARTICLE:27}
\APACinsertmetastar {%
ARTICLE:27}%
\begin{APACrefauthors}%
Vellekoop, M.%
\BCBT {}\ \BBA {} Nieuwenhuis, H.%
\end{APACrefauthors}%
\unskip\
\newblock
\APACrefYearMonthDay{2007}{}{}.
\newblock
{\BBOQ}\APACrefatitle {On Option Pricing Models in the Presence of Heavy Tails}
  {On option pricing models in the presence of heavy tails}.{\BBCQ}
\newblock
\APACjournalVolNumPages{Quantitative Finance}{7}{5}{563-573}.
\PrintBackRefs{\CurrentBib}

\bibitem [\protect \citeauthoryear {%
Zhao%
, Wang%
\BCBL {}\ \BBA {} Song%
}{%
Zhao%
\ \protect \BOthers {.}}{%
{\protect \APACyear {2018}}%
}]{%
ARTICLE:13}
\APACinsertmetastar {%
ARTICLE:13}%
\begin{APACrefauthors}%
Zhao, P.%
, Wang, J.%
\BCBL {}\ \BBA {} Song, Y.%
\end{APACrefauthors}%
\unskip\
\newblock
\APACrefYearMonthDay{2018}{May}{}.
\newblock
{\BBOQ}\APACrefatitle {Optimal Portfolio under Non-Extensive Statistical
  Mechanics and Value-at-Risk Constraints} {Optimal portfolio under
  non-extensive statistical mechanics and value-at-risk constraints}.{\BBCQ}
\newblock
\APACjournalVolNumPages{Acta Physica Polonica A}{133}{5}{1170-1173}.
\PrintBackRefs{\CurrentBib}

\end{thebibliography}
\bibliographystyle{apacite}

\end{document}